\documentclass[iicol]{sn-jnl}

\usepackage{graphics}
\usepackage{graphicx}
\usepackage{color}
\usepackage{hyperref}
\usepackage{bm}
\usepackage{calligra}
\usepackage[T1]{fontenc}
\usepackage{amsmath}
\usepackage{amssymb}
\usepackage{amsfonts}
\usepackage{mathrsfs}  
\usepackage{dsfont}
\usepackage{footmisc}

\usepackage[utf8]{inputenc}

\usepackage{lipsum}
\usepackage{textcomp}
\usepackage{xcolor}

\usepackage{xspace}

\begin{document}

\title[Shifting nucleons]{Methods for systematic study of nuclear structure in high-energy collisions}

\author[1]{Matthew Luzum}
\affil[1]{Instituto de F\'{\i}sica, Universidade de  S\~{a}o Paulo,  Rua  do  Mat\~{a}o, 1371,  Butant\~{a},  05508-090,  S\~{a}o  Paulo,  Brazil}

\author[2]{Mauricio Hippert}
\affil[2]{Illinois Center for Advanced Studies of the Universe\\ Department of Physics, University of Illinois at Urbana-Champaign, 1110 W. Green St., Urbana IL 61801-3080, USA}

\author[3]{Jean-Yves Ollitrault}
\affil[3]{Institut de physique th\'eorique, Universit\'e Paris Saclay, CNRS, CEA, F-91191 Gif-sur-Yvette, France}


\abstract{
There is increasing interest in using high-energy collisions to probe the structure of nuclei, in particular with the high-precision data made possible by collisions performed with pairs of isobaric species.
A systematic study requires a variation of parameters representing nuclear properties such as radius, skin thickness,  angular deformation, and short-range correlations, to determine the sensitivity of the various observables on each of these properties.
In this work we propose a method for efficiently carrying out such study, based on the shifting of positions of nucleons in Monte-Carlo samples.  We show that by using this method, statistical demands can be dramatically reduced --- potentially reducing the required number of simulated events by orders of magnitude --- paving the way for systematic study of nuclear structure in high-energy collisions,
}

\maketitle


\section{Introduction}

An emerging direction of research in recent years is to use data from high-energy nucleus-nucleus collisions in order to infer properties of colliding nuclei, such as their deformation~\cite{Giacalone:2019pca,Giacalone:2020awm,Giacalone:2021udy,Bally:2021qys,Jia:2021qyu,Giacalone:2021uhj,Zhang:2021kxj,Bally:2023dxi,Samanta:2023tom}.
In particular, recent measurements in isobaric collisions of $^{96}$Ru and $^{96}$Zr at the Relativistic Heavy-Ion Collider~\cite{STAR:2021mii} have shown that small differences in the nuclear properties are potentially measurable~\cite{Hammelmann:2019vwd,Xu:2021vpn,Nijs:2021kvn,Zhao:2022uhl,Jia:2022qgl,Liu:2022kvz,Nie:2022gbg}.

For such studies, 
it is desirable to have the ability to continuously change nuclear structure parameters, and precisely quantify the small differences they induce on collision observables. 
Modern modeling of high-energy collisions involves Monte Carlo simulations~\cite{Moreland:2018gsh,JETSCAPE:2020mzn,Nijs:2021kvn,Parkkila:2021yha}.
If a new, independent set of Monte Carlo simulations is generated for each parameter value, a huge amount of statistics is required to resolve the small differences in final observables. 
For example, in Ref.~\cite{Jia:2022qrq}, hundreds of millions of simulation events were generated for each set of possible nuclear parameter values, in order to study the precise observable ratios measured in the RHIC isobar run.

We introduce a method which significantly reduces the required number of simulations, potentially by orders of magnitude. 
The most important source of event-by-event fluctuations in  a high-energy collision is the random position of each nucleon at the time of collision~\cite{PHOBOS:2006dbo,Alver:2010gr,Luzum:2011mm, Luzum:2013yya}. 
We show that changes in nuclear properties can be implemented by slightly shifting these positions. 

Nucleon positions are typically sampled independently according to a Woods-Saxon distribution. 
We first introduce, in Sec.~\ref{s:woodssaxon}, a fast method for carrying out this sampling. 
In Sec.~\ref{s:onebody}, we describe how nucleon positions must be shifted in order to implement a change in the one-body distribution, such as a global deformation. 
In Sec.~\ref{s:twobody}, we describe how they must be shifted in order to implement short-range two-body correlations. 
The application of our method to heavy-ion collisions is discussed in Sec.~\ref{s:hic}. 
In Sec.~\ref{sec:reweighting}  we discuss a complementary method  to explore small regions in parameter space without any extra simulations by reweighting events.
In Sec.~\ref{sec:benchmarks}, we show on specific examples that our methods are considerably more efficient than traditional methods.

\section{Preparing the nucleus}
\label{s:woodssaxon}

The idea is to prepare a set of nuclear configurations --- the position of each nucleon in a nucleus --- each consisting of $A$ independent nucleons governed by a spherically-symmetric probability distribution $\rho$.   This existing set can then be modified by changing the positions of these nucleons, such that the new set of nuclei respect a different 1-body density $\rho$, as well as a non-trivial correlation function $C(r)$.

The typical profile used to describe a nucleus is a Woods-Saxon
\begin{align}
\label{eq:WS}
\rho(r) &= \frac 1 N\frac{1}{1 + e^{\frac {r-R}{a}}},
\end{align}
which is parameterized by a radius $R$ and diffusiveness parameter $a$.  The  proportionality factor $N(R,a)$ is defined by noting that total probability must sum to unity:  $\int d^3x\, \rho = 1$.  Explicitly,
\begin{align}
N(R,a) &= -8 a^3\pi  {\rm Li}_3(-e^{R/a}),
\end{align}
where Li$_3$ is the polylogarithm of order 3.

This distribution can be independently sampled (e.g., using acceptance-rejection sampling ) in order to generate a discrete set of nuclei governed by distribution $\rho$.

Here we present an alternative method for sampling (and manipulating) a spherically-symmetric distribution.  This is done by approximating a Woods-Saxon distribution by the convolution of a spherical step function and a 3D Gaussian.   This has the advantage of giving closed-form analytic expressions, which gives a high degree of analytic control.   For example, 
the relation between the point nucleon density and, e.g., the charge density can be written analytically.   
Additionally, nuclear configurations can be computed without rejection sampling of a Woods-Saxon distribution.   
Finally,  parameters $R, a$ can be easily changed.

The idea is to treat the coordinate of each nucleon as a random variable that, rather than being governed by a Woods-Saxon, is a sum of two other random vectors governed by a spherical step function and spherical Gaussian, respectively.

\begin{align}
P_s({\bf x}) = P_s(r) &= \frac{3}{4\pi R_s^3 }\Theta(R_s-r)\\
P_g({\bf x}) = P_g(r) &= \frac 1 {w^3 \sqrt{8 \pi^3}}  e^{-\frac{r^2}{2 w^2}},
\end{align}
with $r =  \lvert{\bf x}\rvert$.   The position of a nucleon is given by the vector sum of positions drawn from each of these distributions.  The sum of the two random vectors thus follows the convolution
\begin{align}
\rho_c({\bf x}) &= \int  P_s({\bf z}) P_g({\bf x - z}) d^3 z\\
&= 
\frac 3 {8 \pi^{3/2} R_s^3} \biggl[\frac{\sqrt 2 w}{r} \left(e^{-\frac{(r+R_s)^2}{2w^2}} - e^{-\frac{(r-R_s)^2}{2w^2}}  \right) \nonumber \\
&\ \  + \sqrt\pi \left\{{\rm erf}\left(\frac{r + R_s}{\sqrt 2 w}\right)- {\rm erf}\left(\frac{r - R_s}{\sqrt 2 w}\right)\right\}
\biggr]
\label{eq:stepgauss}
\end{align}

\begin{figure}
\includegraphics[width=\linewidth, trim={0 0 0 0},clip]{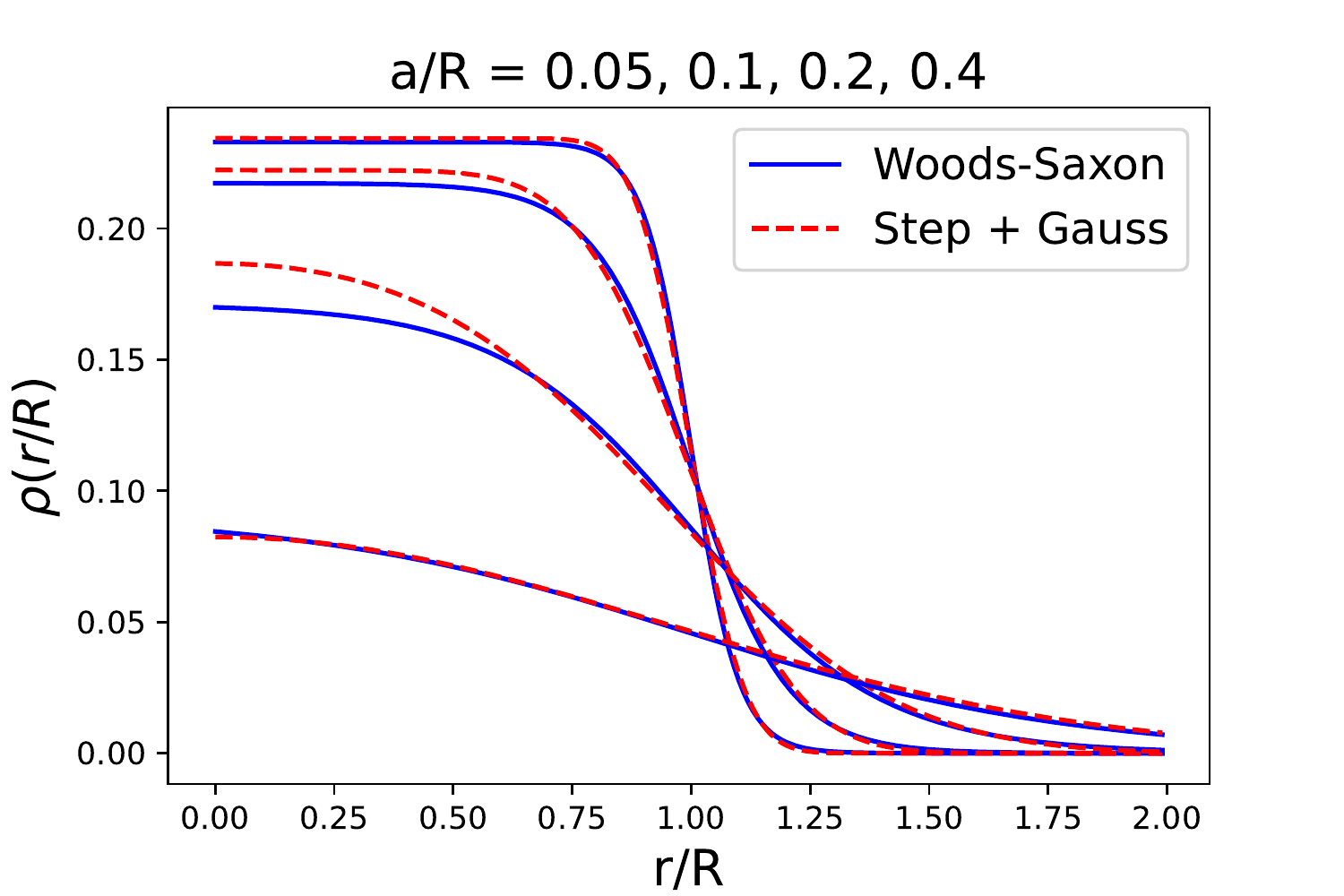}
\caption{
\label{fig:StepCompare}
Comparison of scaled Woods-Saxon distribution $\rho(r/R)$ with step+Gauss distribution $\rho_c(r/R)$, Eq.~\eqref{eq:stepgauss}.  
}
\end{figure}

With the correct choice of radial and width parameters $(R_s, w)$ a Woods-Saxon can be well approximated by this function $\rho_c(r) \simeq \rho(r)$. 

A natural way to compare probability distributions is the Kullback–Leibler (KL) divergence~\cite{Kullback:1951zyt}
\begin{align}
\label{eq:KL}
D_{\rm KL}(\rho_1\vert\vert\rho_2) &\equiv \int d^3 x \rho_1({\bf x}) \log \frac{\rho_1({\bf x})}{\rho_2({\bf x})}.
\end{align}
Two identical distributions have KL divergence  $D_{\rm KL}(\rho_1\vert\vert\rho_1) = 0$, while increasingly different distributions cause it to increase.

 We can choose the best parameters $(R_s, w)$ to describe a given Woods-Saxon $(R,a)$ by minimizing the KL divergence $D_{\rm KL}(\rho, \rho_c)$.    A rough rule of thumb for the conversion when $a/R$ is small is
\begin{align}
R_s(R,a) &\simeq R\left[1+ 1.5 \left(\frac a R\right)^{1.8}\right]\\
w(R,a) &\simeq 1.83\, a
\end{align}
In practice we obtain the correct values on-the-fly with numerical minimization.

A comparison to traditional Woods-Saxon distributions is shown in Fig.~\ref{fig:StepCompare}.  The fit is worst for certain large values of skin thickness $a$ (0.2--0.3 fm), but works well for values of interest.

In all of  the numerical examples shown in this work, we use this method for generating seed nucleon configurations, which can then be modified as desired by shifting nucleon positions.

\section{Changing shape}
\label{s:onebody}

Once a discrete set of nuclear configurations are prepared, they can be modified to follow the desired statistics. 

First we consider a continuous change in the 1-body density --- that is, a change that can be characterized by some continuous parameter $t$, $\rho = \rho({\bf x}, t)$.

We will be transporting particles in order to affect the change in density that is brought about by a change in parameter $t$.  Conservation of particles demands that this transformation obeys a continuity equation
\begin{equation}
  \label{continuity}
 \frac{\partial\rho}{\partial t}  + {\bf\nabla}\cdot\left(\rho\, {\bf v}\right)=0. 
\end{equation} 
where the shift in position $d{\bf x}({\bf x}, t)$ of particles with respect to parameter $t$ is described by  vector field ${\bf v}({\bf x},t)$,
\begin{align}
d{\bf x} = {\bf v} dt.
\end{align}

\subsection{Radial deformation}
We start by describing how to modify the parameters $R, a$ of a spherical Woods-Saxon, in the case where it is desired to start with a given Woods-Saxon sampling instead of the alternate step+Gauss distribution \eqref{eq:stepgauss}.

In this case, there is rotational symmetry both before and after the transformation, and we can choose a purely radial shift for the nucleons.

\begin{align}
\label{radialdef}
{\bf v}(r,\theta,\phi) &= \hat r v_r(r)
\end{align}
The continuity equation \eqref{continuity} becomes
\begin{align}
\frac{\partial \rho}{\partial t} = \frac 1 {r^2} \frac{\partial}{\partial r} \left(r^2 \rho v_r\right),
\end{align}
which we can simply integrate to obtain
\begin{align}
v_r(r,t) &= - \frac{1}{r^2 \rho(r,t)} \int_0^r (r')^2 \frac{\partial\rho}{\partial t}(r',t) dr'
\end{align}

For the specific case of a Woods-Saxon distribution, we note that for fixed value of $a/R$, different distributions are related to each other by a simple scale transformation $ r \to C r$.

To modify $a/R$, we define
\begin{align}
\rho(r,t) &\propto \frac{1}{1 + e^{(r_s- t R_s)}},\\
r_s &= \frac r a\\
R_s &= \frac R a,
\end{align}
so that
\begin{align}
\frac{\partial \rho}{\partial t} &= \frac{R_s}{2} \frac 1 {1 + \cosh(r_s - t R_s)}\\
v_r(r,t) &= - \frac{a}{r^2 \rho} \int_0^{r/a} \frac {(r_s')^2} {1 + \cosh(r'_s - t R_s)} dr_s'
\end{align}

\subsection{Angular deformation}

Starting with a set of spherical nuclei prepared using one of the above methods, we can modify the positions of nucleons to obtain nuclei described by a deformed distribution.  A nucleus with angular deformation is typically parameterized by deforming a Woods-Saxon \eqref{eq:WS}, replacing the radius parameter
\begin{align}
R\to R\left(1 + \sum_{\ell,m} \beta_{\ell,m} Y_{\ell,m}(\theta,\phi)\right),
\end{align}
where $Y_{\ell,m}(\theta,\phi)$ are the real form of spherical harmonics, and each multipole component is characterized by a coefficient $\beta_{\ell,m}$.
That is, 
\begin{align}
\tilde \rho(r, \theta, \phi) &\propto \frac{1}{1 + e^{\frac {r-R  -R\sum_{\ell,m} \beta_{\ell,m} Y_{\ell,m}}{a}}}.
\end{align}

\begin{figure*}
\includegraphics[width=0.5\linewidth, trim={80 0 80 0},clip]{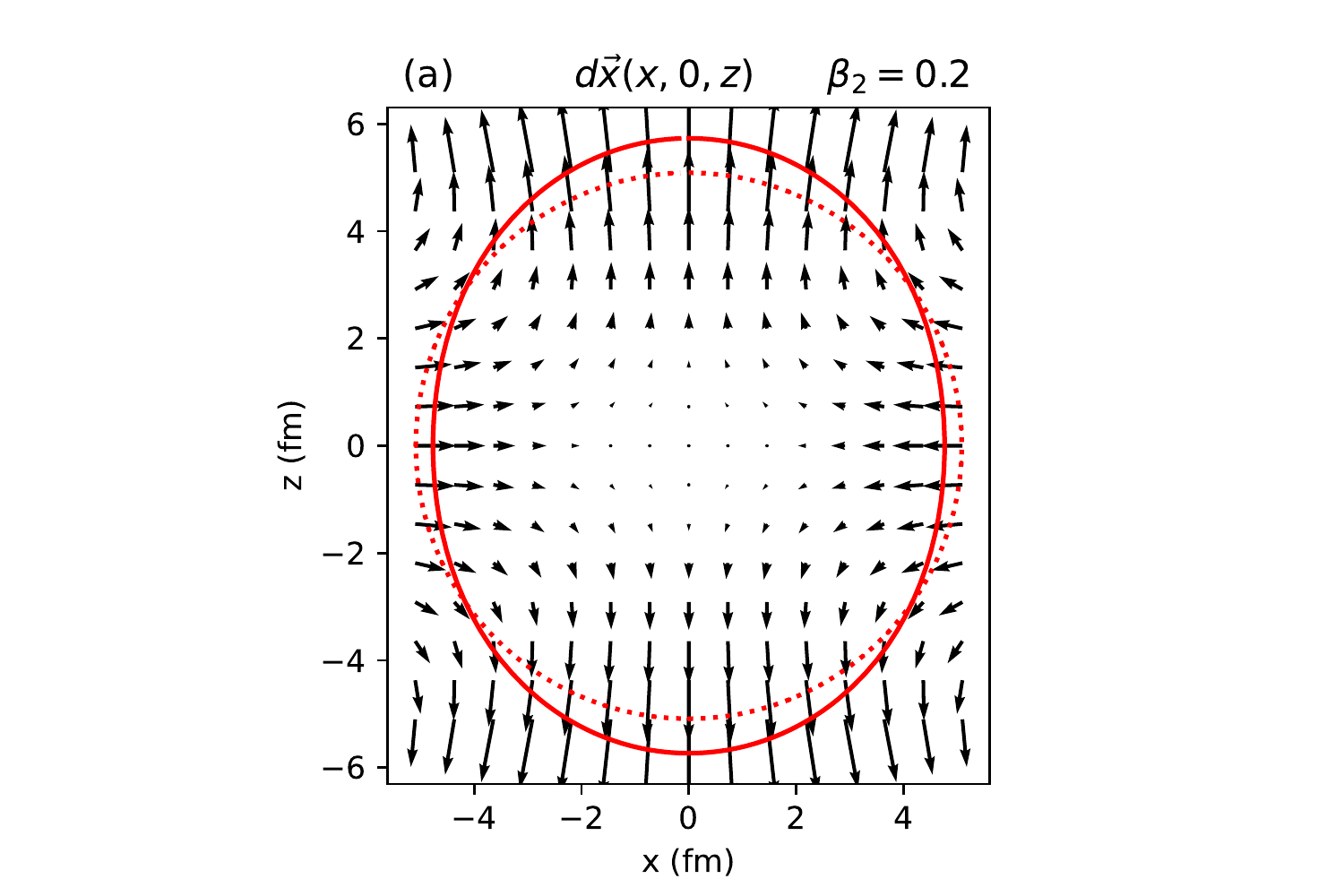}
\includegraphics[width=0.5\linewidth, trim={80 0 80 0},clip]{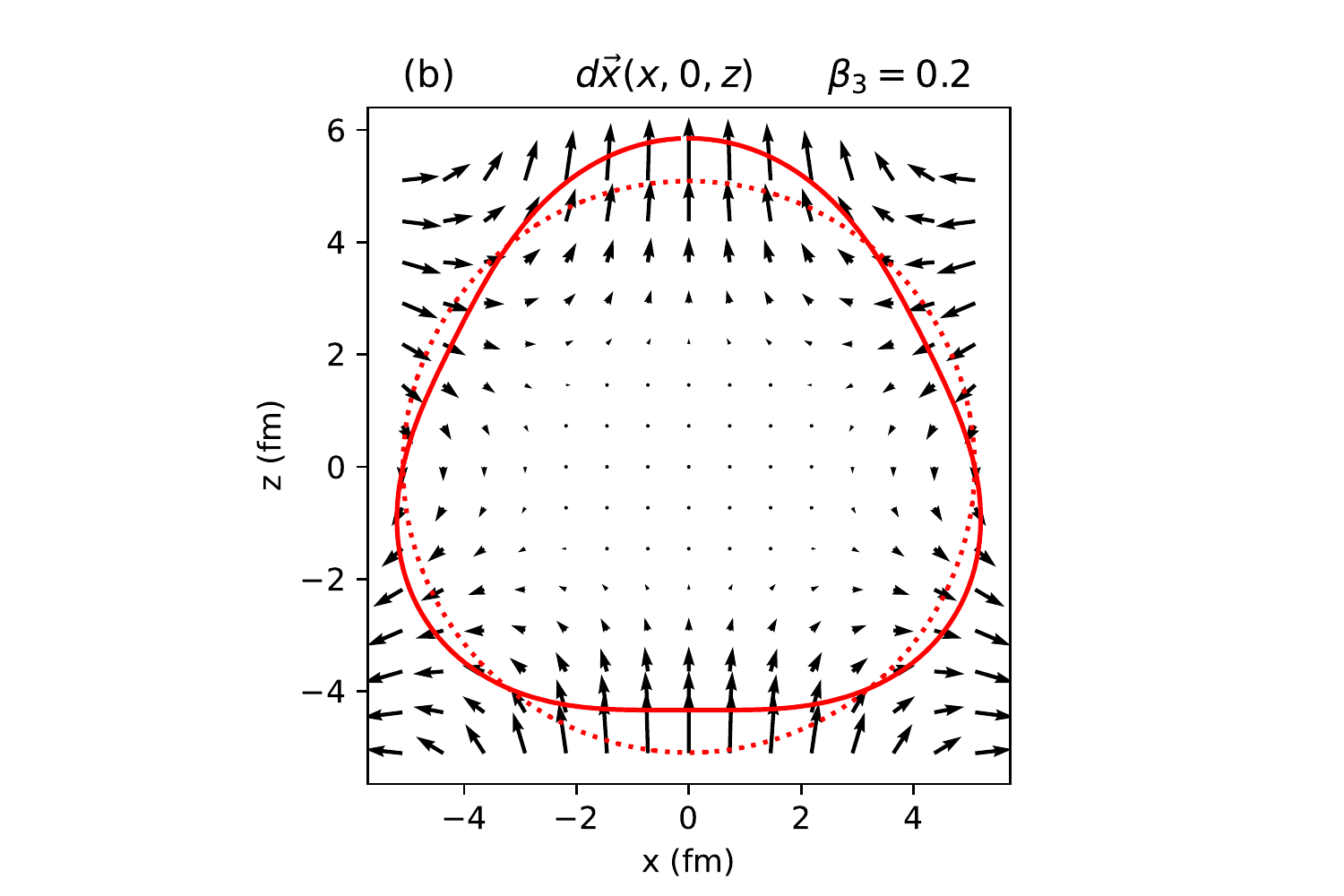}
\caption{
\label{fig:quiver}
Vector plot of shift $d\vec x$ in $x$-$z$ plane for case of axial quadrupole deformation only, $\beta_{2,0} = 0.2 $ (a), and for axial octupole deformation only, $\beta_{3,0} = 0.2$  (b).   
The axis scales correspond to Woods-Saxon radius $R = 5.09$ fm.   
The curves represent the Woods-Saxon radius $R(\theta,\phi)$ for the starting spherical distribution (dotted) and the final deformed distribution (solid).}
\end{figure*}

We first note that we can relate the deformed distributions to the spherical Woods-Saxon \eqref{eq:WS} with a coordinate transformation
\begin{align}
\label{eq:defWS}
\tilde \rho(r, \theta, \phi)  
&\propto
\rho(r - R\sum_{\ell,m} \beta_{\ell,m} Y_{\ell,m}).
\end{align}
We also note that the normalization of the probability distribution depends (weakly) on coefficients $\beta_{\ell,m}$ in addition to parameters $R$, $a$.    However, our transformation will consist only of transporting particles, so probability is naturally conserved. We therefore omit the normalization factor,  which will cancel in the final expressions.

We therefore define our $t$-parameterized distribution as
\begin{align}
\label{tparameterized}
\rho({\bf x}, t) &= \rho(r - t\sum_{\ell,m} R \beta_{\ell,m} Y_{\ell,m}),
\end{align}
so that $t=0$ represents a spherical distribution and $t=1$ represents the desired deformed distribution, with the two limits continuously connected by intermediate values.

We then determine a vector field ${\bf v}$ which satisfies the continuity equation (\ref{continuity}) for the above expression of $\rho({\bf x},t)$.
This single first-order differential equation for the vector field ${\bf v}$ is an underdetermined problem, with no unique solution.  
We posit a solution of the form of the gradient of a scalar field $\Phi$.\footnote{In the language of fluid dynamics, this corresponds to irrotational flow.}:
\begin{equation}
\label{potentialflow}
    {\bf v}={\bf\nabla}\Phi.
\end{equation}
An advantage of this prescription is that symmetry arguments can be used to specify the dependence of $\Phi$ on the coordinates $(r,\theta,\phi)$. 
In the case of a radial deformation, for instance, the velocity field Eq.~(\ref{radialdef}) can be written as a gradient, where $\Phi$ solely depends on the radial coordinate $r$. 

Inserting Eq.~(\ref{potentialflow}) into the continuity equation (\ref{continuity}), one obtains the following equation for $\Phi$:
\begin{equation}
  \label{continuity2}
 \frac{\partial\rho}{\partial t}  + {\bf\nabla}\rho\cdot{\bf\nabla}\Phi+
 \rho\,\Delta\Phi=0. 
\end{equation} 
If the deformation is sufficiently small, we can linearize this equation and replace $\rho$ with the original spherical distribution (corresponding to the value at $t=0$) in the last two terms.
The second term then reduces to $\rho'(r)\partial\Phi/\partial r$.

In the case of an angular deformation, we decompose $\Phi$ into multipole components
\begin{equation}
\Phi(r,\theta,\phi) =   \sum R \beta_{\ell,m} f_{\ell, m}(r) Y_{\ell, m}(\theta,\phi)
\end{equation}
Inserting this equation into Eq.~(\ref{continuity2}) and using Eq.~(\ref{tparameterized}), the equations for the different multipole component decouple, and they satisfy the following second-order equation in $r$
\begin{align}
f_{\ell,m}'' + f_{\ell,m}' \left(\frac 2 r + \frac{\rho'}{\rho}\right) - \frac{\ell(\ell+1)}{r^2}f_{\ell,m} -  \frac{\rho'}{\rho} &= 0.
\end{align}
Note that $f_{\ell,m}$ does not, in fact, depend on $m$.

Defining a unique solution requires two boundary conditions. For this, we consider the behavior at small and large $r$.   Note first that
\begin{align}
\lim_{r\to \infty} \frac{\rho'}{\rho} &= -\frac 1 a\\
\lim_{r\to 0} \frac{\rho'}{\rho} &= \frac{-1}{a \left[1 + e^{\frac{R}{a}} \right]} \sim 0,
\end{align}
since we expect $R\gg a$.

So in the limit of small $r$, $f_{\ell,m}$ is approximately a solution of Laplace's equation
\begin{align*}
\lim_{r\to 0} f_{\ell,m} = C r^\ell + D r^{1-\ell}
\end{align*}

At large $r$, we have
\begin{align}
f_{\ell,m}'' + \frac{1}{a} f'_{\ell,m} - \frac 1 a  &= 0,
\end{align}
with general solution
\begin{align}
\lim_{r\to \infty} f_{\ell,m}  = r + C' e^{\frac r a} + D'.
\end{align}
We would like to choose boundary conditions which eliminate both the divergence at small $r$, that is, $D=0$ and the divergence at large $r$, that is, $C'=0$.    This can be achieved with conditions
\begin{align}
f_{\ell,m}(r\to 0) &= 0\\
f'_{\ell,m}(r\to\infty) &= 1
\end{align}

\begin{table}[t]
\begin{center}
\begin{tabular}{ l  | c c c c c }
 System  & $R$ (fm) & $a$ (fm) & $\beta_2$ & $\gamma$ ($^\circ$) & $\beta_{3,0}$ \\ 
 \hline
 $^{96}$Ru & 5.09 & 0.46 & 0.16 &  30 & 0\\  
 $^{96}$Zr & 5.02 & 0.52 & 0.06 &  0 & 0.20\\   
\end{tabular}
\end{center}
\caption{Sample parameter set relevant for studying the isobar systems of $^{96}$ Ru and $^{96}$Zr of Ref.~\cite{STAR:2021mii}.  The resulting densities are illustrated in Fig.~\ref{fig:density}.
\label{tab:1}
}
\end{table}

\begin{figure*}
\includegraphics[height=4.7cm, trim={0 0 0 0},clip]{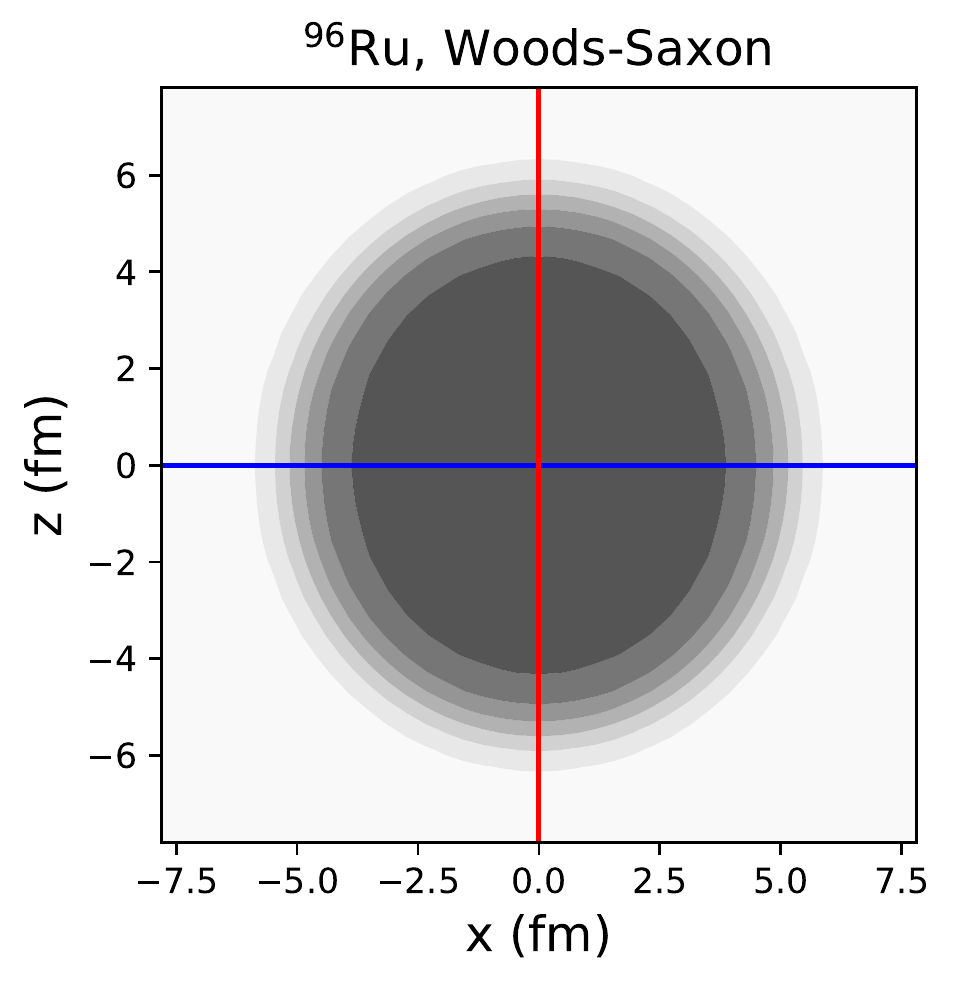}
\includegraphics[height=4.7cm, trim={0 0 0 0},clip]{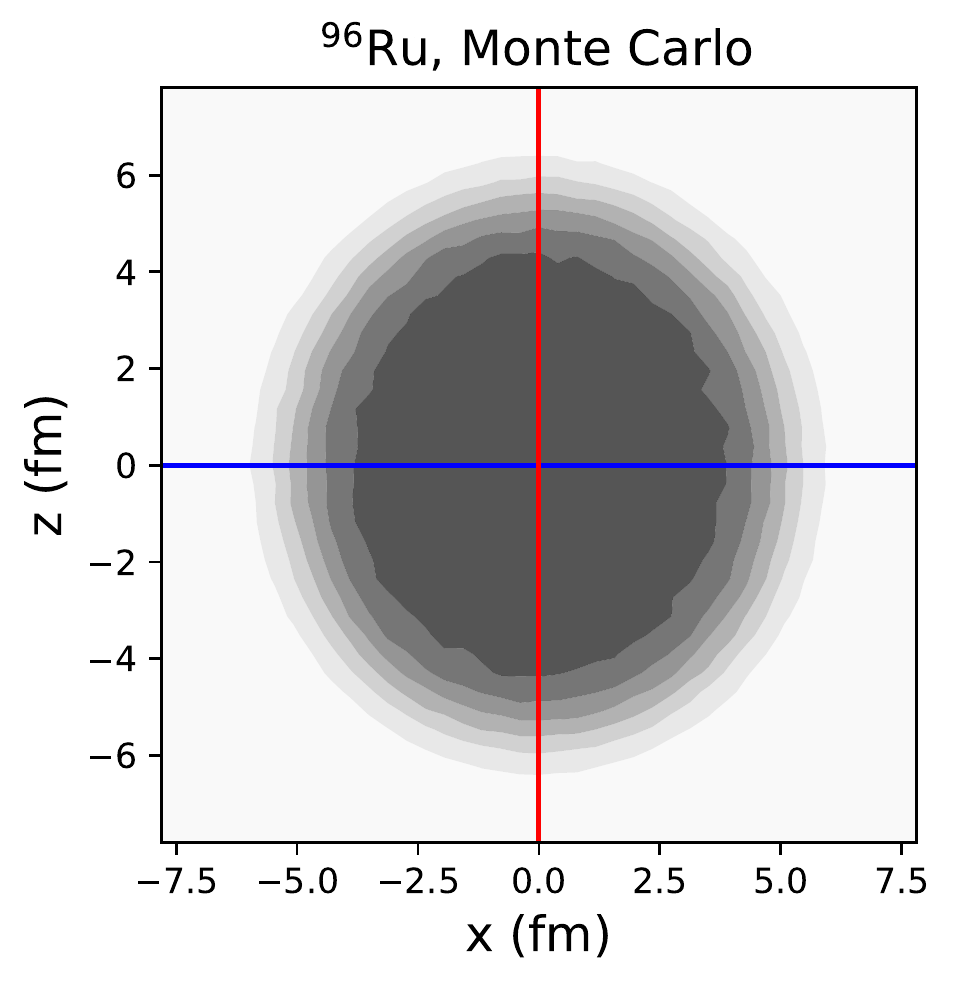}
\includegraphics[height=4.7cm, trim={0 0 0 0},clip]{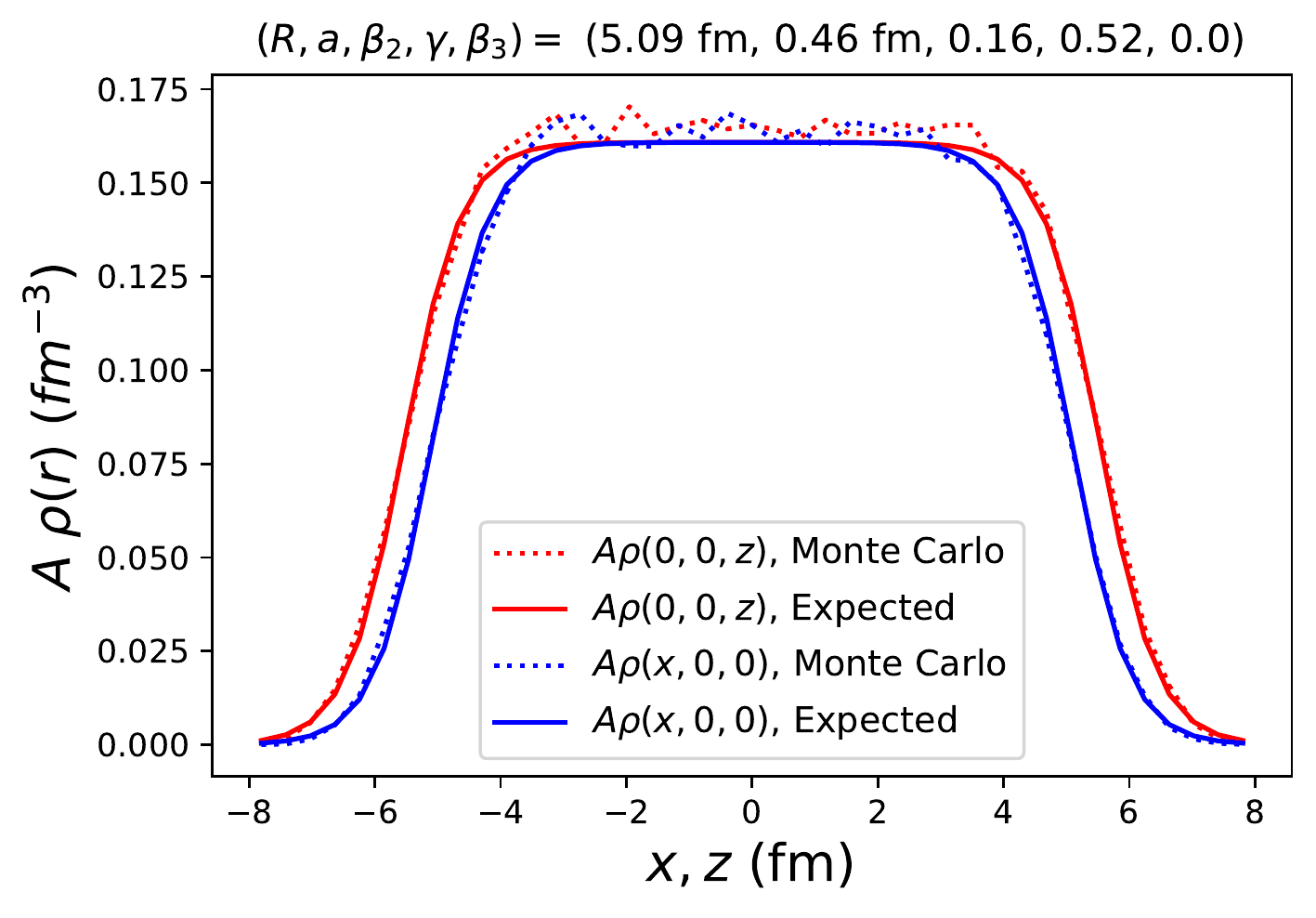}\\
\includegraphics[height=4.7cm, trim={0 0 0 0},clip]{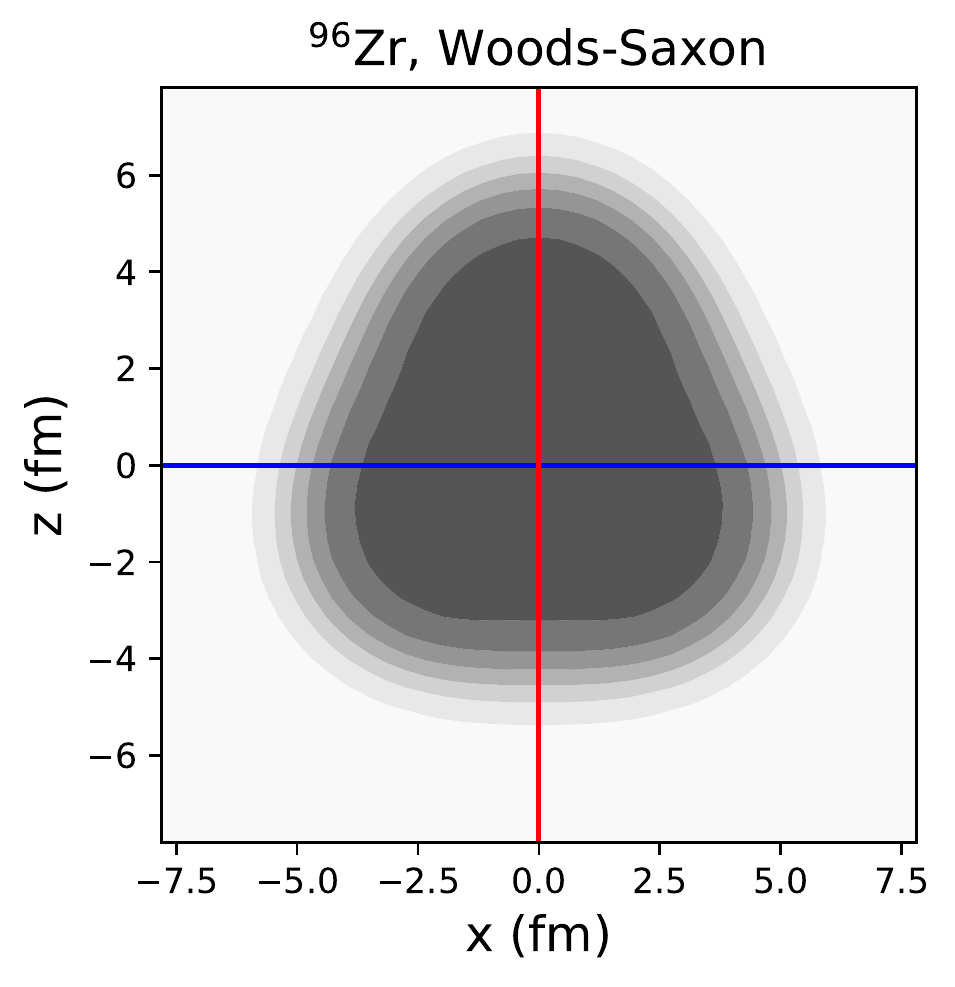}
\includegraphics[height=4.7cm, trim={0 0 0 0},clip]{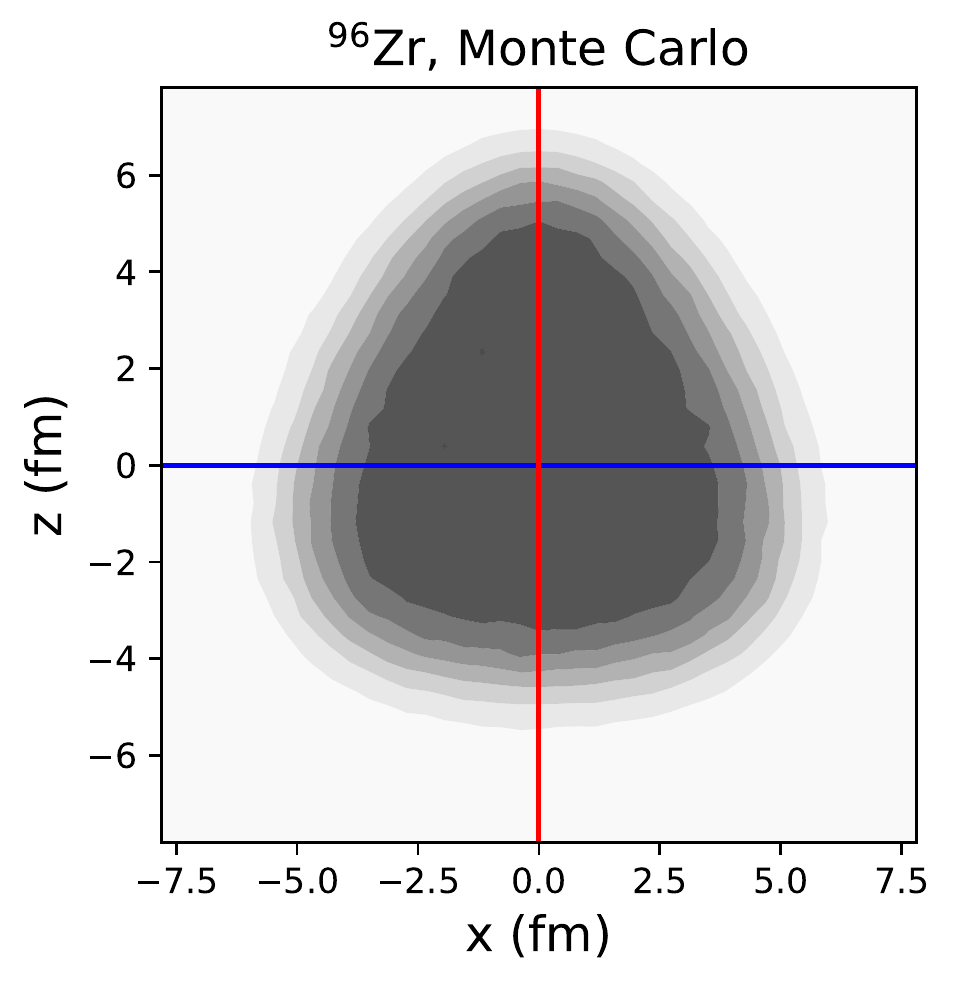}
\includegraphics[height=4.7cm, trim={0 0 0 0},clip]{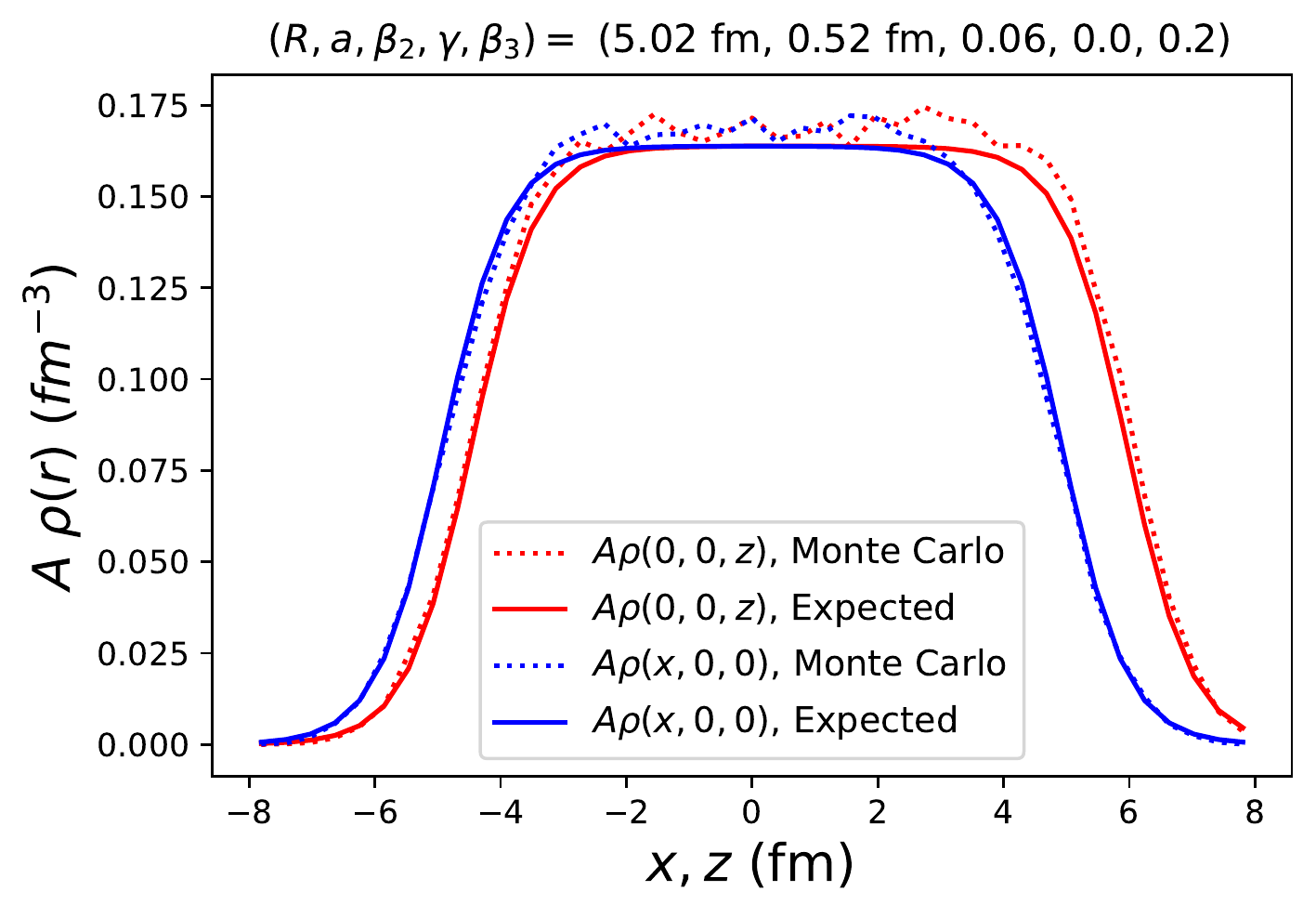}
\caption{
\label{fig:density}
Numerical density $\rho({\bf x})$ from a Monte Carlo sampling versus the expected Woods-Saxon distribution for parameters relevant to $^{96}$Ru (top) and $^{96}$Zr (bottom) as listed in Table\ref{tab:1}.  The Monte Carlo density was obtained by sampling 400,000 nuclei of 96 nucleons each from a spherical (step + Gauss) distribution, performing the deformation, and computing the average density in cubic bins of size $(0.8\ {\rm fm})^3$.   On the left are shown contour plots of the density at $y=0$ in the $x$-$z$ plane, and on the right is the density  along the central $x$ (red) and $z$ (blue) axes
}
\end{figure*}

The general solution can not be found analytically, and must instead be solved numerically.
In practice, this can be done by separately solving the full inhomogeneous and the homogeneous equations, integrating backward from some large $r_{\rm max}$, and taking the linear combination of the two solutions that satisfies the desired boundary condition at some small $r_{\rm min}$.  That is, we solve

\begin{align}
f_I'' + f_I' \left(\frac 2 r + \frac{\rho'}{\rho}\right) - \frac{\ell(\ell+1)}{r^2}f_I -  \frac{\rho'}{\rho}&=0\\
f_I(r_{\rm max}) &= r_{\rm max}\\
f'_I(r_{\rm max}) &= 1\\
f_H'' + f_H' \left(\frac 2 r + \frac{\rho'}{\rho}\right) - \frac{\ell(\ell+1)}{r^2}f_H &= 0  \\
f_H(r_{\rm max}) &= 1\\
f'_H(r_{\rm max}) &= 0
\end{align}
and take as the final solution the linear combination
\begin{align}
f_{\ell,m}(r) &= f_I(r) - \frac{f'_{I}(r_{\rm min})}{f'_{H}(r_{\rm min})} f_H (r).
\end{align}
The solution is very stable and typically does not have any significant dependence on $r_{\rm min}$ or $r_{\rm max}$. 

The final shift is thus
\begin{align}
\label{angletrans}
d{\bf x} &= t \nabla \Phi = t R \sum \beta_{\ell,m} \nabla f_{\ell, m}(r) Y_{\ell, m}(\theta,\phi)\\
&= \hat r \left[t R \sum_{\ell,m} \beta_{\ell,m}  f'_{\ell,m} Y_{\ell,m}\right]\nonumber\\
&\qquad + \hat \theta \left[t \frac {R} r \sum_{\ell,m} \beta_{\ell,m} f_{\ell,m}  \frac{\partial}{\partial \theta} Y_{\ell,m}\right]\nonumber \\
&\qquad +  \hat \phi \left[t \frac {R} {r\sin\theta} \sum_{\ell,m} \beta_{\ell,m}  f_{\ell,m} \frac{\partial}{\partial \phi}  Y_{\ell,m}\right]
\end{align}
For small deformation, we can set $t=1$, shift all particles, and the new set of configurations will consist of independent nucleons with (approximately) the desired 1-body probability density.
For illustration, in Fig.~\ref{fig:quiver} we show example vector plots of the  shift field $d{\bf x}$ for the case of $\beta_{2,0}$  only and $\beta_{3,0}$ only.

For larger deformations ($\sum \beta \gtrsim 0.2$), it is advantageous to break into multiple ($N$) steps of size $t/N$, evaluating the shift $d{\rm x}$ at the new position after each step. 
In principle this requires solving the differential equation for ${\bf v}$ at the new nonzero value of $t$ for each subsequent step.  
In this case, the equations for different multipoles $f_{\ell,m}$ become coupled to each other and it is necessary to simultaneously solve the coupled differential equations. 
In practice,  good accuracy can be obtained by breaking the shift into several steps, but reusing the same ($t=0$) shift field to compute each step.   


For illustration, in Fig.~\ref{fig:density} we explicitly show the resulting density from the Monte Carlo implementation in the relevant case of the isobar pair $^{96}$Ru and $^{96}$Zr from Ref.~\cite{STAR:2021mii} using parameters shown in Table \ref{tab:1}.   Note that quadrupole deformations are typically parameterized with magnitude $\beta_2$ and angle $\gamma$, related to the spherical harmonic coefficients as
\begin{align}
\beta_{2,0} &= \beta_2 \cos(\gamma)\\
\beta_{2,2} &= \frac {1} {\sqrt 2}\beta_2 \sin(\gamma).
\end{align}

\section{Short-range correlations}
\label{s:twobody}

Interactions between nucleons can induce correlations.   Such correlations are encoded in $N$-body distributions beyond the 1-body density $\rho{(\bf x)}$. Let $\rho_2({\bf x}_1, {\bf x}_2)$ denote the distribution of nucleon pairs, from which we can define the correlation function $C({\bf x}_1, {\bf x}_2)$,
\begin{equation}
\label{defC}
  \rho_2({\bf x}_1,{\bf x}_2)=\rho({\bf x}_1)\rho({\bf x}_2)\left[1+C({\bf x}_1,{\bf x}_2)\right]. 
\end{equation}

In particular, short-range interactions between nucleons (as well as their Fermi-Dirac statistics) are expected to induce short-range correlations between nuclei.  That is, you will find fewer (or more) pairs of nucleons within a short distance $r$ from each other than would be expected from independent particles governed by distribution $\rho$.

As was possible for the 1-body distribution, we can induce a correlation between particles by slightly shifting their positions.   The simplest way is to start with a collection of uncorrelated nucleons (prepared, e.g., as described in the previous sections), and change the distance between pairs of nucleons such that the resulting distribution obeys the desired correlation function.

We choose a prescription such that pairs of uncorrelated particles with relative position ${\bf r = } r\hat r = {\bf x}_2 - {\bf x}_1$ are shifted radially from their center point so that their new separation is $\tilde r(r)$.   The new separation is chosen as a monotonically-increasing function that preserves the number of pairs,

\begin{align}
    \int_0^r dr' r'^2 &= \int_0^{\tilde r} dr' r'^2 \left[
    1+C({\bf x}_1,{\bf x}_2)
    \right].
\end{align}
For a chosen correlation function $C$ (which in principle could depend on absolute position, orientation, spin, etc.), one can invert this expression to solve for $\tilde r$.  

We choose a symmetric shift, so that each particle in the pair moves half the required separation shift.
The total shift of a given particle $i$ is the vector sum of the shifts 
implied by its pairing with all the other particles in the nucleus,
\begin{align}
\label{eq:corrshift}
    d{\bf x}_{i} &=  \sum_{j\neq i} \frac 1 2\left(\tilde r_{ij} - r_{ij}\right)  
    \hat r_{ij}.
\end{align}

\begin{figure}
    \centering
    \includegraphics[width=0.7\linewidth]{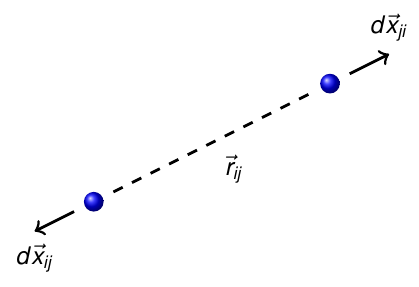}
    \caption{Each particle in a pair is moved radial away from (or toward) their midpoint, in order to induce the desired pair correlation.}
    \label{fig:Corr_Shift}
\end{figure}

Further, we note that the correlation cannot be an arbitrary function.  
In particular, the number of nucleons (and therefore the number of pairs) in each nucleus is fixed, and this places a restriction on the correlation function.  
In our formalism, this corresponds to the fact that $\rho_2$ and $\rho$ are probabilities, which therefore integrate to unity, and which implies the  sum rule
\begin{equation}
\label{sumrule}
 \int \rho({\bf x}_1)\rho({\bf x}_2)C({\bf x}_1,{\bf x}_2)d^3x_1d^3x_2=0. 
\end{equation} 
In order to implement an arbitrary short-range correlation $C_{\rm short}({\bf r})$ , we add a small, constant offset $C_\infty$ to ensure that this sum rule is satisfied (if it's not already), that is:
\begin{equation}
\label{offset}
    C({\bf r}) = C_{\rm short}({\bf r}) + C_{\infty}.
\end{equation}
Inserting this equation into Eq.~(\ref{sumrule}) and assuming that the range of the correlation is much smaller than the system size, one obtains: 
\begin{equation}
\label{cinfty}
C_{\infty} \simeq C_{\rm vol} 
\int d^3 x \rho({\bf x})^2,
\end{equation} 
where $C_{\rm vol}$ is the volume integral of the short-range correlation:
\begin{equation}
\label{cvol}
C_{\rm vol} \equiv\int d^3 r C_{\rm short}({\bf r}).
\end{equation}

\begin{figure}
    \centering
    \includegraphics[width = 0.6\linewidth]{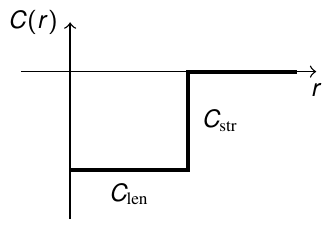}
    \caption{Simple step-function correlation function used in our tests.}
    \label{fig:my_label}
\end{figure}

To illustrate the method, we choose a simple example of a step-function correlation function with a variable correlation length $C_{\rm len}$ and strength $C_{\rm str}$ (see Fig.~\ref{fig:my_label}):
That is, we define:
\begin{align}
    C_{\rm short} &=
    \begin{cases}
        C_{\rm str},  & r \leq C_{\rm len}\\
        0, & r > C_{\rm len}.
    \end{cases}
\end{align}
Eq.~(\ref{cvol}) then gives
\begin{equation}
    C_{\rm vol} = C_{\rm str} \frac 4 3 \pi C_{\rm len}^3.
\end{equation}

We note that for the short-range correlations of interest, the offset $C_{\infty}$ defined by Eq.~(\ref{cinfty}) is quite small.  For example, for $C_{\rm str} = -1$ and $C_{\rm len} = 0.4 $ fm, we obtain $C_{\infty}\simeq 3\times 10^{-3}$, which is seemingly negligible.  Nevertheless, its inclusion ensures that the 1-body distribution $\rho(\vec x)$ is more precisely maintained.

With this choice, the final pair separation ${\bf \tilde r}$ for a given initial separation ${\bf r}$ is

\begin{align}
    \tilde {\bf r}({\bf r}) &=
    \begin{cases}
        {\bf r} \left(
        1 + C_{\rm str} + C_\infty
        \right)^{-1/3} & r \leq r_{\rm sw} \\
        {\bf r} \left(
        \frac{1 - C_{\rm str} C_{\rm len}^3/r^3}{1 + C_\infty}
        \right)^{1/3} & r > r_{\rm sw}
    \end{cases}\\
    r_{\rm sw} &= C_{\rm len}(1 +  C_{\rm str} + C_\infty)^{1/3}.
\end{align}

\begin{figure}
\includegraphics[width=\linewidth, trim={0 0 0 0},clip]{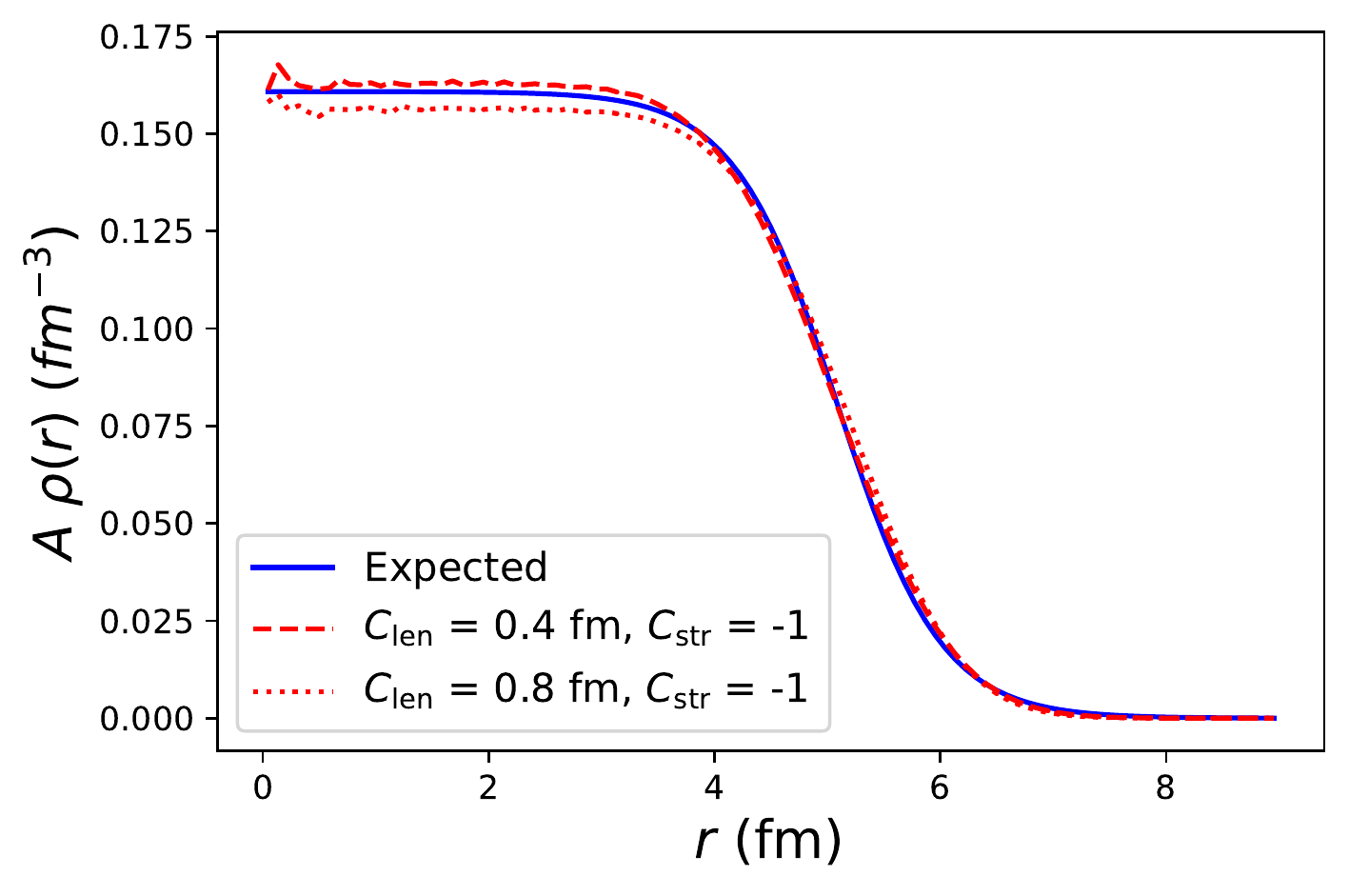}
\caption{
\label{fig:corrdensity}
Average nucleon density ($A\ \rho$) after implementation of step-function 2-body correlation with $C_{\rm str} = -1$ and $C_{\rm len} = 0.4$ fm (dashed) and 0.8 fm (dotted) via particle shift, compared to the expected Woods-Saxon.  1-body Woods-Saxon parameters correspond to the $^{96}$Ru from Table \ref{tab:1}.
}
\end{figure}

\begin{figure}
\includegraphics[width=\linewidth, trim={0 0 0 0},clip]{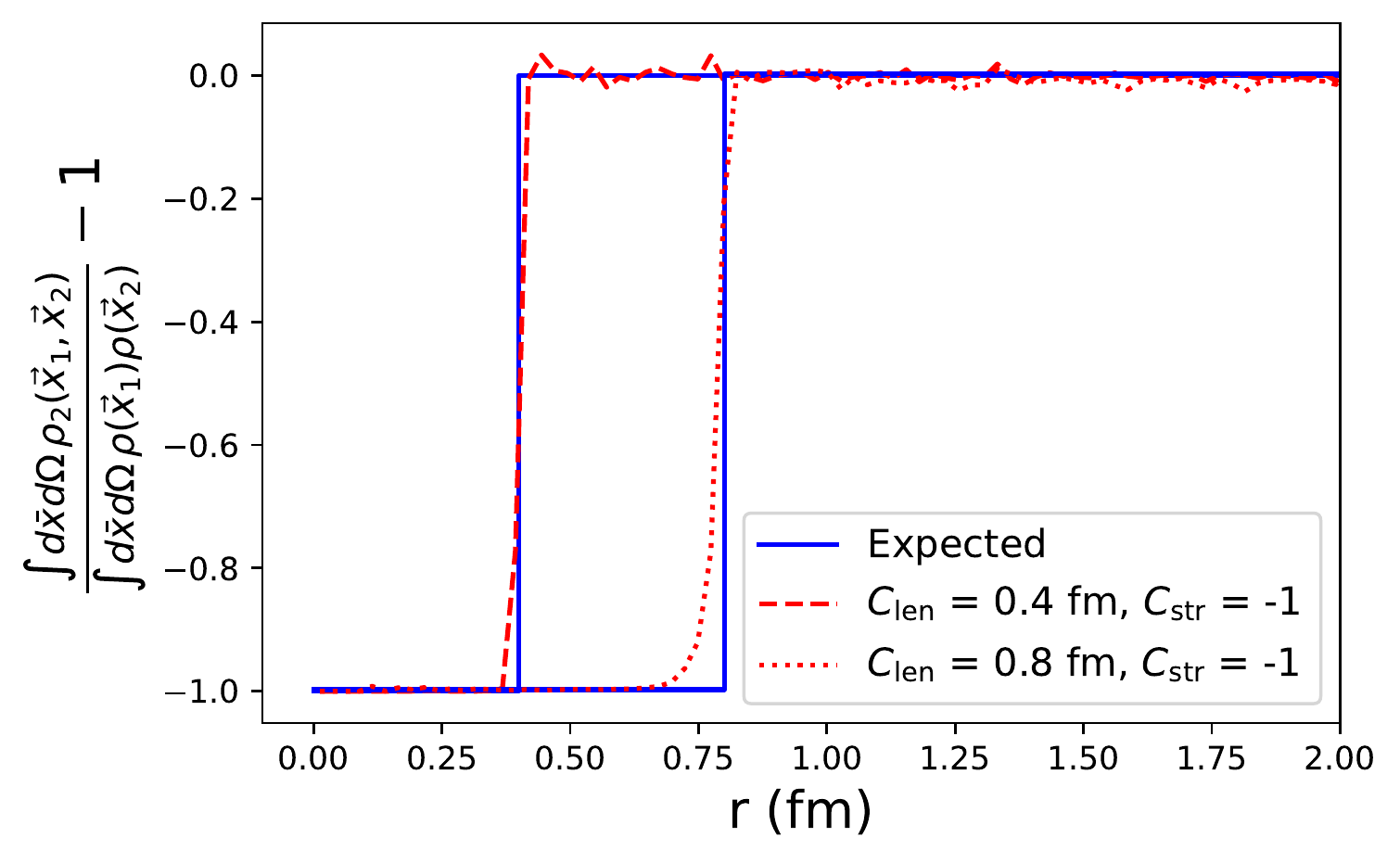}
\caption{
\label{fig:correlationfunction}
Correlation function computed from an independent Monte Carlo sampling plus particle shift, compared to expected pair distribution.
}
\end{figure}

While this method does not guarantee that the 1-body distribution will remain fixed, as long as one inputs a valid correlation function that respects the sum rule \eqref{sumrule}, both the 1-body and 2-body distributions are reproduced to a good approximation for correlations of short range.  See, e.g., Figs \ref{fig:corrdensity} and \ref{fig:correlationfunction}.

Indeed, there are advantages to this method of implementing correlations compared to other common methods, even beyond the increase of computation efficiency that is the main concern of this work.  The excellent simultaneous control of the average density and correlations is one example.  A common simplified method for mimicking a short-range nucleon-nucleon correlation is to simply disallow nuclei with nucleon pairs less than some exclusion distance $d_{\rm min}$.   Besides being restricted to a very specific correlation function, this method can bring other problems.  Implementing this in the most straightforward way \cite{Alver:2008aq}, for example, results in an unwanted modification of the 1-body distribution if $d_{\rm min}$ is not sufficiently small.    

One can use clever methods to keep the nucleon density fixed while ensuring that no pairs have distance less than $d_{\min}$, as done in Ref.~\cite{Moreland:2014oya}.  However this results in an uncontrolled 2-body distribution, which can develop complicated and unintended structures.  Further, the method cannot be used for nuclear shapes without an axial symmetry (for example, for $\beta_{22} \neq 0$), since it involves randomly reassigning the azimuthal angle of offending nucleons.

The sampling method of Refs.~\cite{Alvioli:2009ab,Alvioli:2010yk}, on the other hand, can in principle give more precise control over the 1-body and 2-body distributions.  However, it remains difficult to efficiently and systematically study changes in the nuclear correlations, where the method described here has the clear advantage.

\section{Application to high-energy nuclear collisions}
\label{s:hic}

Once the nuclear configurations are prepared, they can be used to simulate collision events.  The details depend on the specific model used for the collision dynamics, but generally, in order to minimize statistical uncertainty, it is useful to reuse as many of the fluctuations as possible.   Besides the nucleon-position fluctuation discussed in the previous sections, there are others that can arise.

Fluctuations that will be present in any simulation include the random impact parameter of each collision and the orientation of each nucleus that collides.   One should ensure that when preparing collisions corresponding to different nuclear parameters, one uses the same pair of nuclei prepared as described above, with the same impact parameter, and the same Euler angles for each respective nucleus.   

Note that this will not always be possible.   For example, at a given impact parameter, one pair of nuclei may suffer a collision while a corresponding pair of deformed nuclei might not.   To avoid bias, one must nevertheless include such unpaired events, despite the  increase in statistical uncertainty.   Similarly for models with a notion of participating versus spectator nucleons, there might be a different number of participants in corresponding collisions. 

Despite this, the decrease in statistical uncertainty from using our methods can still be dramatic, as we show in Sec.~\ref{sec:benchmarks}.  

The treatment of any other fluctuations will be specific to each model.  Generally, one should reuse fluctuations as much as possible, using the same random number associated with each pair of nuclei (in the case of probabilistic cross section), individual nucleons, transverse position, etc., even if the relevant cumulative probability distribution is slightly different.

\section{Probability Reweighting}
\label{sec:reweighting}
Instead of performing separate simulations for each nucleus (i.e., each point in parameter space), it can be possible to study multiple parameter values from a single set of simulations.   

The idea is the following.   If we sample nuclei randomly from some $A$-body probability distribution $\rho_A({\bf x}_1, {\bf x}_2, \ldots {\bf x}_A)$ we can reinterpret the resulting set of nucleon configurations as a weighted sampling from a different distribution $\rho'_A$.   The relative weight is simply $\rho'/\rho$ evaluated at the sampled nucleon positions.

In the limit of infinite sampling, $\rho'$ can be reconstructed exactly, as long as the distributions both have non-zero support in the same region.  Thus, in principle one can explore many points in parameter space ``for free'', by computing a set of simulations at one point in parameter space (i.e., from a single $A$-body density), and computing results for other parameter sets by simply reweighting the events when computing observables.  

However, convergence can be very poor if the probability distributions are not similar --- especially if $\rho$ has negligible density in some region where $\tilde \rho$ has non-negligible density.  It can take many, many samples to fill out the distribution, and convergence can be extremely poor, wiping out any potential efficiency gains.

Here we make a preliminary study of this method to gauge its feasibility in the context of the nuclear structure studies of interest for this work.  
We use the $A$-body density
\begin{align}
    \rho_A({\bf x}_1, \ldots {\bf x}_A) &= \prod_{i=1}^A \tilde\rho({\bf x}_i)
    \prod_{i, j>i}^A \left[1 - C({\bf x}_i, {\bf x}_j)\right],
\end{align}
with $\tilde\rho$ the (potentially) deformed Woods-Saxon \eqref{eq:defWS} and $C$ a desired 2-particle correlation.  We sample one distribution (an uncorrellated, spherical Woods-Saxon), and estimate observables relevant to a different distribution by reweighting events constructed from the sampled nuclei.

Details and numerical benchmarks are presented in Section \ref{sec:benchmarks}.

\section{Quantifying the efficiency gain}
\label{sec:benchmarks}

In order to quantify the efficacy of these methods, we implement a simple Glauber model and calculate the ratio of observables obtained from collisions of nuclei with two different nuclear properties.

For simplicity, we set impact parameter to 0, so that each nucleus (before sampling) is centered at the same transverse coordinate.   After generating a set of nuclear configurations, we select two nuclei to collide and give each nucleus a random (3D) orientation.

A nucleon-nucleon collision is said to occur if the transverse distance is $r < 2 \sqrt{\sigma_{\rm NN}/\pi}$, where $\sigma_{NN}$ = 0.62 fm$^2$ represents the inelastic nucleon-nucleon cross section.

Among the set of participating nucleons, we then calculate the participant eccentricities
\begin{align}
\varepsilon_n &= \frac{\lvert \sum_i r_i^n e^{in\phi_i} \rvert}{\sum_i r_i^n}
\end{align}
for $n=2$ and $n=3$, where $(r_i, \phi_i)$ are the polar coordinates of participant nucleon $i$ in the transverse plane.

With the eccentricity computed in each collision event, we compute the RMS eccentricity
\begin{align*}
\varepsilon_n\{2\} &\equiv \sqrt{\langle \varepsilon_n^2 \rangle}
\end{align*}

The ratio of $\varepsilon_n\{2\}$ in two different collision systems is a good approximation for the ratio of final measured observables $v_n\{2\}$, and so represents a quantity of high relevance.

To estimate the uncertainty in the eccentricity ratio, we use jackknife resampling.

We quantify the efficacy of our method by comparing the statistical uncertainty for the case of traditionally-prepared nuclei --- that is, where all nuclei are sampled completely independently --- to the case where we use the method of shifted nucleons to relate different nuclear structure.  In the latter case, we use the same seed nuclei (before shift) for both nuclei in the colliding pair, and the same rotation angle for each corresponding nucleus.  

For these tests we start with a baseline spherical nucleus of uncorrelated nucleons, with Woods-Saxon parameters $R = 5.09$ fm and $a = 0.46$ fm.  We then compare this baseline case to the case where one parameter ($\beta_2$, $\beta_3$, $C_{\rm len}$) is non-zero.  For correlations we study a (near) full exclusion, $C_{\rm str} = -1$, but note that preiliminary tests indicate that eccentricities only depend on the combination $C_{\rm str} C_{\rm len}^3$.

The deformed/correlated nuclei are either generated completely independently or by shifting nucleons.   Observables are then computed for these cases, as well as the case where the baseline spherical, uncorrelated collisions are reweighted to compute eccentricity ratios corresponding to the respective deformed/correlated cases.

\begin{figure}
\includegraphics[width=\linewidth, trim={0 0 0 0},clip]{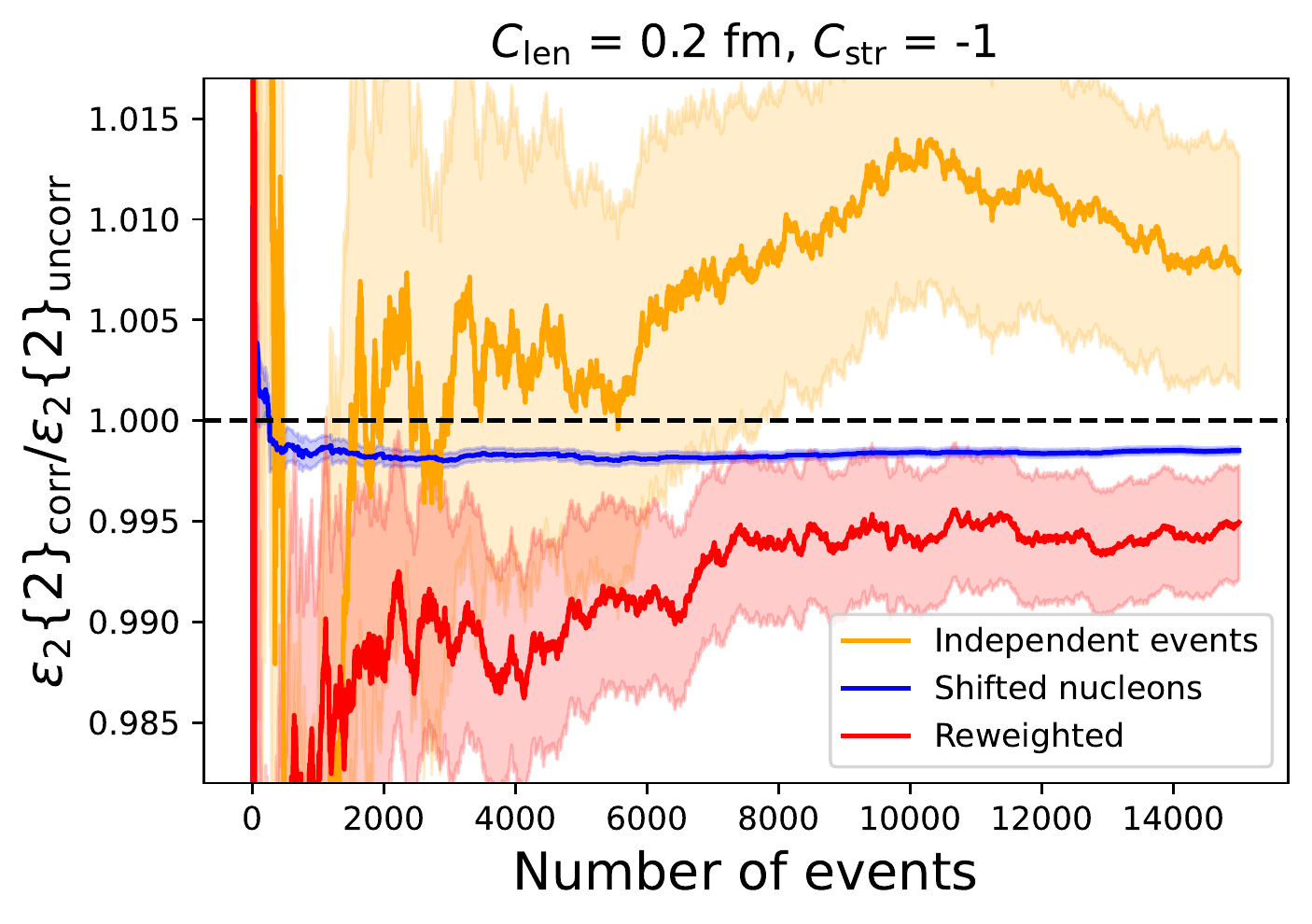}
\caption{Comparison of  observable ratio $\epsilon_2\{2\}$ for nuclei with short-range correlations (numerator) and uncorrelated nucleons (denominator), for case of step-function correlation with length $C_{\rm len}$ = 0.2 fm and strength $C_{\rm str}$ = -1.   Using the method of shifting nucleon positions, the statistical uncertainty is decreased by a factor $\sim$32 compared to the traditional independent sampling, so that a calculation with 35 events has smaller uncertainty than 100000 events sampled independently.  Reweighting the uncorrelated events to obtain correlated nuclei gives a smaller uncertainty than independent sampling of the correlated nuclei, but is not competitive with the shifting-nucleons method in this case.
\label{fig:e2ratio}
}
\end{figure}

In Fig~\ref{fig:e2ratio}, we show an example of adding short range correlations ($C_{\rm length}$ = 0.2 fm, $C_{\rm strength}$ = -1, numerator of eccentricity ratio) to the case where nucleons are uncorrelated.   The shift in eccentricity with the addition of correlations is $\sim$0.1\%, which requires a very large number of collision events to resolve using the traditional method.  In contrast, using the method of shifting nucleons, the statistical uncertainty is reduced by a factor $\sim$32.
The required number of events to obtain a certain statistical precision is thus reduced by a factor $\sim$2900.   That is, a calculation with only $\sim$35 events has smaller uncertainty than 100000 events sampled independently.  Such a dramatic decrease in computation requirements has clear implications on the feasibility of systematic study.

However, this efficiency gain depends on the size of the transformation between the two nuclei.   In particular, as the  typical shift in nucleon position increases in distance, there is a higher likelihood of participant nucleons turning to spectators, or vice versa.  This diminishes the correlation in statistical error that exists before and after the transformation.    

In order to give an idea of the efficiency gain in various contexts, we list a number of examples in Table \ref{tab:2}.  

The change in $\varepsilon_n\{2\}$ generally tends to scale linearly with $\beta^2$ and $C_{\rm str} C_{\rm len}^3$.  The improvement factor tends to scale like 1/$\sqrt{\Delta(\beta^2)}$ and $1/\Delta(C_{\rm str}C_{\rm len}^3)$.  That is, while smaller changes create greater difficulty for traditional independent sampling, when using the better method uncertainties are actually \textit{decreased}.   

For a given average nucleon shift, the study of correlations sees more efficiency gains than the study of angular deformation.  This is because in the case of angular deformations, nucleons in the bulk of the nucleus do not shift, while nucleons near the edge shift \textit{more} than the average (see, e.g., Fig.~\ref{fig:quiver}).  It is precisely these nucleons at the edges that have the largest effect on eccentricities (especially in the central collisions tested here), and so the loss of syncronization from participant-spectator conversion due to the nucleon shift causes a decrease in correlation of statistical uncertainty, and therefore a decrease in efficiency gain.

The method of reweighting improves even more quickly with decreasing changes in nuclear properties than the shifting-nucleons method.  So it is more efficient than shifting nucleons for small changes ($\Delta\beta \lesssim $ 0.01), but loses efficacy quickly for larger changes, becoming worse than independent sampling for $\Delta\beta \gtrsim 0.08$ or $\Delta(C_{\rm str} C_{\rm len}^3) \gtrsim$ (0.3 fm$)^3$ and rapidly degrading beyond that.    Thus, even though the same set of events can be used in principle to study many points in parameter space, the range of parameter space that can be efficiently explored is limited.  So it can not be used as a general replacement for the shifting-nucleons method unless one wants only to explore a quite small parameter space.  However, it may be very useful in conjunction with other methods.  For example, larger jumps in parameter space obtained from shifting nucleons can be filled out with reweighting around each point in parameter space that is sampled and simulated.   In that way, one can obtain a more precise interpolation between sampled parameter values, without computing any extra simulations.

\begin{table}[t]
\begin{center}
\begin{tabular}{ c c c c c c}
         & Param. &  $\varepsilon_n\{2\}$ & Improv. & Avg.\\ 
 Par.  & Change &  Change & Factor & Shift\\
 \hline
$(\beta_2)^2$ & $(0.005)^2$ &  0.02\% & 170 & 0.008 fm\\ 
$(\beta_2)^2$ & $(0.01)^2$ & 0.10\% & 100 & 0.02 fm\\ 
$(\beta_2)^2$ & $(0.02)^2$ &  0.39\% & 42 & 0.03 fm\\  
$(\beta_2)^2$ & $(0.05)^2$ &  2.3\% & 12 & 0.08 fm\\  
$(\beta_2)^2$ & $(0.1)^2$ &  8.8\% & 4.7 & 0.17 fm\\  
$(\beta_2)^2$ & $(0.2)^2$ &  31\% & 2.1 & 0.33 fm\\  
$(\beta_3)^2$ & $(0.01)^2$ &  0.05\% & 79 & 0.01 fm\\  
$(\beta_3)^2$ & $(0.05)^2$ &  1.6\% & 13 & 0.06 fm\\  
$(\beta_3)^2$ & $(0.1)^2$ &  6.3\% & 5.0 & 0.12 fm\\  
$(\beta_3)^2$ & $(0.2)^2$ &  23\% & 2.2 & 0.25 fm\\  
$(C_{\rm len})^3$ & (0.2 fm$)^3$ &  0.13\% & 2900  & 0.002 fm\\   
$(C_{\rm len})^3$ & $\times 2$ &  0.27\% & 1100 & 0.005 fm \\   
$(C_{\rm len})^3$ & $\times 4$ &  0.53\% & 350 & 0.009 fm \\   
$(C_{\rm len})^3$ & (0.4 fm$)^3$ &  1.1\% & 180  & 0.017 fm\\   
$(C_{\rm len})^3$ & $\times 2$ &  2.0\% & 98  & 0.032 fm\\   
$(C_{\rm len})^3$ & $\times 4$ &  3.8\% & 54  & 0.059 fm\\   
$(C_{\rm len})^3$ & (0.8 fm$)^3$ &  7.3\% & 25 & 0.11 fm \\   
$(C_{\rm len})^3$ & $\times 2$ &  14\% & 13 & 0.19 fm \\   
\end{tabular}
\end{center}
\caption{
Reduction in the required number of events (``Improv.~Factor'') for computing the ratio of RMS eccentricity $\varepsilon_n\{2\}$ using the method of shifting nucleons compared to independent nuclei, in a simple benchmark test of $b=0$ collisions in a participant Glauber model.  Nuclei were prepared from an uncorrelated, spherical distribution and modified so that a single parameter becomes non-zero.  Also shown is the corresponding change in $\varepsilon_n\{2\}$ and the average nucleon shift when changing each parameter from a value of 0.  For the $\beta_3$ change, column 3 represents the resulting change in $\varepsilon_3\{2\}$, otherwise it is $\varepsilon_2\{2\}$.  The improvement factor does not change significantly with harmonic, so only column 3 is harmonic-dependent.   For the case of correlated nucleons, the strength parameter is set to $C_{\rm str} = -1$, corresponding to almost complete exclusion within the correlation length.
\label{tab:2}
}
\end{table}

\section{Conclusions}

We have introduced methods to dramatically decrease statistical demands when studying how heavy-ion collision observables depend on the properties of the colliding nucleons.  These properties include the average nucleon density (parameterized by a Woods-Saxon radius $R$ and diffusiveness $a$ along with any number of angular deformation coefficients $\beta_{\ell,m}$) as well as an arbitrary short-range correlation function $C({\bf x}_1, {\bf x}_2)$. 

In general, the efficiency gain depends on the specific context, as well as the physical model used for simulations, with the largest benefit corresponding to the study of small changes in nuclear structure.  Nevertheless, these methods always reduces the statistical requirements, and quite dramatic improvements are possible, with some cases seeing a reduction of the necessary number of simulations by multiple orders of magnitude.

Armed with these techniques, a large number of detailed and systematic studies of nuclear structure in the context of relativistic heavy-ion collisions will be possible.

\section*{Acknowledgments}
We thank Govert Nijs and Dean Lee for suggesting to try a reweighting method.
We thank the Institute for Nuclear Theory at the University of Washington for its hospitality and the Department of Energy for partial support during the completion of this work.
ML acknowledges the support of the São Paulo Research Foundation (FAPESP) under grants 2021/08465-9, 2018/24720-6, and 2017/05685-2, as well as the support of the Brazilian National Council for Scientific and Technological Development (CNPq). 
We thank support from the ``Emilie du Ch\^atelet'' visitor programme and from the GLUODYNAMICS project  funded by the “P2IO LabEx (ANR-10-LABX-0038)” in the framework “Investissements d’Avenir” (ANR-11-IDEX-0003-01) managed by the Agence Nationale de la Recherche (ANR), France.
MH was supported 
by the National Science Foundation (NSF) within the
framework of the MUSES collaboration, under grant number OAC-2103680.


\begin{thebibliography}{99}

\bibitem{Giacalone:2019pca}
G.~Giacalone,
Phys. Rev. Lett. \textbf{124}, no.20, 202301 (2020)
doi:10.1103/PhysRevLett.124.202301
[arXiv:1910.04673 [nucl-th]].

\bibitem{Giacalone:2020awm}
G.~Giacalone,
Phys. Rev. C \textbf{102}, no.2, 024901 (2020)
doi:10.1103/PhysRevC.102.024901
[arXiv:2004.14463 [nucl-th]].

\bibitem{Giacalone:2021udy}
G.~Giacalone, J.~Jia and C.~Zhang,
Phys. Rev. Lett. \textbf{127}, no.24, 242301 (2021)
doi:10.1103/PhysRevLett.127.242301
[arXiv:2105.01638 [nucl-th]].

\bibitem{Bally:2021qys}
B.~Bally, M.~Bender, G.~Giacalone and V.~Som\`a,
Phys. Rev. Lett. \textbf{128}, no.8, 082301 (2022)
doi:10.1103/PhysRevLett.128.082301
[arXiv:2108.09578 [nucl-th]].

\bibitem{Jia:2021qyu}
J.~Jia,
Phys. Rev. C \textbf{105}, no.4, 044905 (2022)
doi:10.1103/PhysRevC.105.044905
[arXiv:2109.00604 [nucl-th]].

\bibitem{Giacalone:2021uhj}
G.~Giacalone, J.~Jia and V.~Som\`a,
Phys. Rev. C \textbf{104}, no.4, L041903 (2021)
doi:10.1103/PhysRevC.104.L041903
[arXiv:2102.08158 [nucl-th]].

\bibitem{Zhang:2021kxj}
C.~Zhang and J.~Jia,
Phys. Rev. Lett. \textbf{128}, no.2, 022301 (2022)
doi:10.1103/PhysRevLett.128.022301
[arXiv:2109.01631 [nucl-th]].

\bibitem{Bally:2023dxi}
B.~Bally, G.~Giacalone and M.~Bender,
[arXiv:2301.02420 [nucl-th]].

\bibitem{Samanta:2023tom}
R.~Samanta and P.~Bozek,
[arXiv:2301.10659 [nucl-th]].

\bibitem{STAR:2021mii}
M.~Abdallah \textit{et al.} [STAR],
Phys. Rev. C \textbf{105}, no.1, 014901 (2022)
doi:10.1103/PhysRevC.105.014901
[arXiv:2109.00131 [nucl-ex]].

\bibitem{Hammelmann:2019vwd}
J.~Hammelmann, A.~Soto-Ontoso, M.~Alvioli, H.~Elfner and M.~Strikman,
Phys. Rev. C \textbf{101}, no.6, 061901 (2020)
doi:10.1103/PhysRevC.101.061901
[arXiv:1908.10231 [nucl-th]].

\bibitem{Xu:2021vpn}
H.~j.~Xu, H.~Li, X.~Wang, C.~Shen and F.~Wang,
Phys. Lett. B \textbf{819}, 136453 (2021)
doi:10.1016/j.physletb.2021.136453
[arXiv:2103.05595 [nucl-th]].

\bibitem{Nijs:2021kvn}
G.~Nijs and W.~van der Schee,
[arXiv:2112.13771 [nucl-th]].

\bibitem{Zhao:2022uhl}
S.~Zhao, H.~j.~Xu, Y.~X.~Liu and H.~Song,
[arXiv:2204.02387 [nucl-th]].

\bibitem{Jia:2022qgl}
J.~Jia, G.~Giacalone and C.~Zhang,
[arXiv:2206.10449 [nucl-th]].

\bibitem{Liu:2022kvz}
L.~M.~Liu, C.~J.~Zhang, J.~Zhou, J.~Xu, J.~Jia and G.~X.~Peng,
Phys. Lett. B \textbf{834}, 137441 (2022)
doi:10.1016/j.physletb.2022.137441
[arXiv:2203.09924 [nucl-th]].

\bibitem{Nie:2022gbg}
M.~Nie, C.~Zhang, Z.~Chen, L.~Yi and J.~Jia,
[arXiv:2208.05416 [nucl-th]].

\bibitem{Moreland:2018gsh}
J.~S.~Moreland, J.~E.~Bernhard and S.~A.~Bass,
Phys. Rev. C \textbf{101}, no.2, 024911 (2020)
doi:10.1103/PhysRevC.101.024911
[arXiv:1808.02106 [nucl-th]].

\bibitem{JETSCAPE:2020mzn}
D.~Everett \textit{et al.} [JETSCAPE],
Phys. Rev. C \textbf{103}, no.5, 054904 (2021)
doi:10.1103/PhysRevC.103.054904
[arXiv:2011.01430 [hep-ph]].

\bibitem{Parkkila:2021yha}
J.~E.~Parkkila, A.~Onnerstad, S.~F.~Taghavi, C.~Mordasini, A.~Bilandzic, M.~Virta and D.~J.~Kim,
Phys. Lett. B \textbf{835}, 137485 (2022)
doi:10.1016/j.physletb.2022.137485
[arXiv:2111.08145 [hep-ph]].

\bibitem{Jia:2022qrq}
J.~Jia, G.~Giacalone and C.~Zhang,
[arXiv:2206.07184 [nucl-th]].

\bibitem{PHOBOS:2006dbo}
B.~Alver \textit{et al.} [PHOBOS],
Phys. Rev. Lett. \textbf{98}, 242302 (2007)
doi:10.1103/PhysRevLett.98.242302
[arXiv:nucl-ex/0610037 [nucl-ex]].

\bibitem{Alver:2010gr}
B.~Alver and G.~Roland,
Phys. Rev. C \textbf{81}, 054905 (2010)
[erratum: Phys. Rev. C \textbf{82}, 039903 (2010)]
doi:10.1103/PhysRevC.82.039903
[arXiv:1003.0194 [nucl-th]].

\bibitem{Luzum:2011mm}
M.~Luzum,
J. Phys. G \textbf{38}, 124026 (2011)
doi:10.1088/0954-3899/38/12/124026
[arXiv:1107.0592 [nucl-th]].

\bibitem{Luzum:2013yya}
M.~Luzum and H.~Petersen,
J. Phys. G \textbf{41}, 063102 (2014)
doi:10.1088/0954-3899/41/6/063102
[arXiv:1312.5503 [nucl-th]].

\bibitem{Kullback:1951zyt}
S.~Kullback and R.~A.~Leibler,
The Annals of Mathematical Statistics \textbf{22}, no.1, 79-86 (1951)
doi:10.1214/aoms/1177729694

\bibitem{Alver:2008aq}
B.~Alver, M.~Baker, C.~Loizides and P.~Steinberg,
[arXiv:0805.4411 [nucl-ex]].

\bibitem{Moreland:2014oya}
J.~S.~Moreland, J.~E.~Bernhard and S.~A.~Bass,
Phys. Rev. C \textbf{92}, no.1, 011901 (2015)
doi:10.1103/PhysRevC.92.011901
[arXiv:1412.4708 [nucl-th]].

\bibitem{Alvioli:2009ab}
M.~Alvioli, H.~J.~Drescher and M.~Strikman,
Phys. Lett. B \textbf{680}, 225-230 (2009)
doi:10.1016/j.physletb.2009.08.067
[arXiv:0905.2670 [nucl-th]].

\bibitem{Alvioli:2010yk}
M.~Alvioli and M.~Strikman,
Phys. Rev. C \textbf{83}, 044905 (2011)
doi:10.1103/PhysRevC.83.044905
[arXiv:1008.2328 [nucl-th]].
\end{thebibliography}
\end{document}